\documentstyle[12pt,epsf]{article}

\catcode`@=11
\def\simle{\mathrel{\mathpalette\@versim<}}   
\def\simge{\mathrel{\mathpalette\@versim>}}   
\def\@versim#1#2{\lower2.5pt\vbox{\baselineskip0pt \lineskip-.5pt
   \ialign{$\m@th#1\hfil##\hfil$\crcr#2\crcr\sim\crcr}}}
\catcode`@=12

\begin{document}
\begin{center}
\bf
Monte Carlo Algorithm for the Double Exchange Model\\
Optimized for Parallel Computations
\end{center}

\begin{center}
  Nobuo Furukawa

{\sl Department of Physics, Aoyama Gakuin University,\\
 Setagaya, Tokyo 157-8572, Japan}\\

\vspace{5mm}
 Yukitoshi Motome

{\sl Institute for Materials Science, Tsukuba University,\\
Tsukuba Japan}\\

\vspace{5mm}
 Hisaho Nakata

{\sl Scientific and Engineering Research Center,
 Aoyama Gakuin University,\\
 Setagaya, Tokyo 157-8572, Japan}\\

\end{center}

\vspace{5mm}

A new algorithm  for Monte Carlo calculation 
of the double exchange model is studied.
The algorithm is commonly applicable to  wide
classes of strongly correlated
electron systems which involve itinerant electrons
coupled with thermodynamically fluctuating fields.
Using moment expansions of the density of states with
Chebyshev polynomials,
the algorithm provides an efficient calculation on 
large size clusters, especially on parallel computers.
Benchmark calculations are performed
on Beowulf-type  cluster
systems with over 100 CPUs in parallel.


\vspace{1cm}


\pagebreak

\section{Introduction}

Double exchange model \cite{Furukawa99-} has been
widely studied as a model for itinerant ferromagnetism,
which is one of the most interesting subject
in the  field of strong electron correlations.
The model was introduced by Zener in 1951 
 as a canonical model for perovskite manganites \cite{Zener51},
where  $e_{\rm g}$ and $t_{2\rm g}$ electrons of manganese $3d$ bands
 are treated as itinerant electrons and localized spins, respectively.
Interactions between these two species of electrons
are taken into account  through on-site Hund's couplings
which are stronger than electron hopping energies.
The model has been investigated extensively for half a century,
especially  in connection with the 
 colossal magnetoresistance phenomena in manganites.

Although the model is simple in its form,  thermodynamic
properties have not been well known so far.
In general, it is  difficult to investigate
strongly correlated electron systems, both in 
analytical and numerical methods.
Especially in this model, effects of thermal fluctuations 
around the critical temperature are quite strong
so that perturbational approaches
as well as mean-field methods do not work in controlled manners.

Recently, dynamical mean-field methods
as well as Monte Carlo (MC) calculations on small lattice clusters
are performed \cite{Furukawa99-}.
These results are more reliable than simple mean-field theories
in the sense that they partially take into account
the thermal fluctuations. Nevertheless, the former method
completely neglects spatial fluctuations while the latter 
suffers from finite size errors due to the loss of
long wavelength fluctuations.

In order to study thermodynamic properties of the model, especially
critical phenomena and their influences to electron conductions,
it is necessary to perform calculations which properly
take into account fluctuation effects.
MC calculation on a large size cluster 
seems to be a promising method, provided finite-size
extrapolation are treated properly. 

In this paper, we study an algorithm for 
a MC calculation of the double exchange model
which improves CPU-time consumption significantly.
Using this algorithm, calculations for large size systems
become easier, which enables us to perform extrapolations
to the thermodynamic limit as well as finite-size scalings.

\section{Monte Carlo algorithm}

The double exchange model is defined by
\begin{equation}
 {\cal H}(\{ \vec S_i \}) 
= -t \sum_{<i,j>\sigma} c_{i\sigma}^\dagger c_{j\sigma} - 
J_{\rm H} \sum_i \vec \sigma_i \cdot \vec S_i,
\end{equation}
where $c$ and $c^\dagger$ represent operators for itinerant electrons
while $\vec S$ is the localized classical spins.
On-site Hund's coupling $J_{\rm H}$ gives the 
interaction energy
 between itinerant  electrons and localized spins.
For a given and fixed
 spin configuration  $\{ \vec S_i \}$, the Hamiltonian
is equivalent to a single-body electron system interacting
with random magnetic fields.
Properties of the system at finite temperatures
are obtained through the thermodynamic average over 
the configuration  $\{ \vec S_i \}$.

Boltzmann weight for a spin configuration $\{ \vec S_i \}$ is given by
\begin{equation}
P(\{ \vec S_i \}) = 
   {\rm Tr} \exp[ -\beta ( {\cal H}(\{ \vec S_i \}) - \mu N)]
=    \prod_m  \left( \vphantom {\prod}
	   1 + \exp[ -\beta (\varepsilon_m - \mu)]
        \right),
    \label{GCtrace}
\end{equation}
where Tr represents a grand canonical trace over
fermion degrees of freedom.
Here, $\varepsilon_m $ represent eigenvalues
for ${\cal H}(\{ \vec S_i \}) $.
MC sampling of the
spin configuration $\{ \vec S_i \} $ with
the probability density $P(\{ \vec S_i \})$ 
gives stochastical estimates for
thermodynamical properties of the system.
Direct evaluation of eq.~(\ref{GCtrace}) requires 
all the eigenvalues of ${\cal H}(\{ \vec S_i \}) $.
To obtain all of them  through a full
diagonalization of the Hamiltonian matrix,
it is necessary to make a calculation of $O(N_{\rm dim}^3)$ where
$N_{\rm dim}$ is the Hilbert space dimension of the Hamiltonian.

Alternatively, 
 $P(\{ \vec S_i \}) $ can be obtained
using the density of states (DOS) \cite{Motome99}. 
Equation~(\ref{GCtrace}) is rewritten as
\begin{equation}
\log P(\{ \vec S_i \}) = 
  \int \!{\rm d}\varepsilon D(\varepsilon)
\log\! \left( \vphantom{\sum}
       1 + \exp[-\beta( \varepsilon -\mu)]   \right),
  \label{GCdos}
\end{equation}
where  $D(\varepsilon)$ is the DOS which depends on
the spin configuration $\{ \vec S_i\}$.
Equation~(\ref{GCdos}) is calculated efficiently by
the moment expansion algorithm \cite{Wang94,Silver94}. 
For a given function $f(\varepsilon)$, we perform
a Chebyshev polynomial expansion
\begin{eqnarray}
&& \int {\rm d}\varepsilon D(\varepsilon) f(\varepsilon)
 =  \mu_0 f_0 + 2\sum_{n\ge 1}  \mu_n f_n ,\\
&& f_n = \int \frac{{\rm d}\varepsilon }{\pi\sqrt{1-\varepsilon^2}}
T_n(\varepsilon) f(\varepsilon), \\
&& \mu_n = {\rm Tr} T_n({\cal H}) 
  = \sum_{\nu} \langle \nu | T_n({\cal H}) | \nu \rangle,
  \label{Trace}
\end{eqnarray}
where  $T_n$ is the $n$-th Chebyshev polynomial defined by
$T_0(\varepsilon)=1$, $T_1(\varepsilon)=\varepsilon$, and
$ T_{m+1}(\varepsilon) = 2 \varepsilon T_m(\varepsilon) - T_{m-1}
(\varepsilon)$. $\{|\nu\rangle \}$ is a complete set of kets.
We assume here that,
in order to ensure the expansion, 
 the Hamiltonian is normalized properly to satisfy $|\varepsilon_m |<1$.
Moments for the DOS $\mu_n$ are calculated for each given spin
configuration $ \{ \vec S_i \} $, while $f_n$ are fixed
throughout the MC run.

In practice, eq.~(\ref{Trace}) is calculated as follows.
We define
\begin{equation}
  |\nu ;m\rangle \equiv T_m ({\cal H})|\nu\rangle .
  \label{Def-nu-m}
\end{equation}
Using the recursion formula for the Chebyshev polynomials,
we have
\begin{equation}
  |\nu; m\rangle = 2{\cal H} |\nu; m-1\rangle  - |\nu; m-2\rangle.
\end{equation}
From the multiplication formula $T_{m+n} = 2 T_m T_n - T_{m-n}$, 
moments for the DOS are calculated as
\begin{eqnarray}
  \mu_{2m} &=& \sum_\nu \left( \vphantom{\sum}
   \langle \nu;m | \nu;m \rangle -1
\right) ,
  \label{mu-even}
\\
  \mu_{2m+1} &=& \sum_\nu \left( \vphantom{\sum}
   \langle \nu;m | \nu;m+1 \rangle -\langle \nu;0 | \nu;1 \rangle 
\right).
  \label{mu-odd}
\end{eqnarray}

As we see from above,
calculation of $ P(\{ \vec S_i \}) $ based on the
moment expansion algorithm
costs the CPU time of $O(N_{\rm dim}^2)$, 
{\em i.e.} the sparse matrix product $\sim O(N_{\rm dim})$ times
the trace over the complete set $\sim O( N_{\rm dim})$,
provided the electron hopping is short ranged.
Therefore, 
when
the lattice size is large,
this algorithm provides us a
faster calculation than the algorithm based on matrix diagonalization 
with $O(N_{\rm dim}^3)$.
Moreover, for each $\nu$,
 eqs.~(\ref{Def-nu-m})-(\ref{mu-odd})
can be calculated independently. Thus this algorithm 
is optimized for parallel computations. In contrast,
the previous algorithm based on diagonalization of matrices
is known to be inefficient for parallelizations.

This algorithm is  applicable to  wide
classes of strongly correlated
electron systems where
itinerant electrons are
coupled with thermally fluctuating fields $\{ \phi_i\}$.
If one considers $\{ \phi_i\}$ as a classical field,
thermodynamic averages over $\{ \phi_i\}$ are calculated
by a MC run. $P(\{ \phi_i\})$ are calculated
by the moment expansion algorithm.
It is justified to treat $\{ \phi_i\}$ as a classical field at finite
temperature, at least 
near the renormalized classical critical points.
It is in general interesting to survey  such a region,
since  the effects
of  critical fluctuations to the conduction electrons
are highly  non-trivial.

\section{Benchmark Results and Comments}

Numerical calculations are
 performed on {\em Aoyama Plus} systems \cite{AoyamaPlus-},
Beowulf-type
clusters of commodity personal computers
connected by 100Base-Tx Fast Ethernet.
Benchmarks are taken on two of the cluster systems,
(i) 69 node cluster of dual Pentium II 350MHz with
384MB memory (total 138 processors and 26GB memory) and
(ii) 11 node cluster of dual Celeron 533MHz with 256MB memory
(total 22 processors and 2.8GB memory).
We use MPI for parallel computations.
Within a node, shared memory SMP communications 
based on OpenMP interfaces are used.

\begin{table}[htb]
\begin{center}
 \begin{tabular}{ccc}
 \hline
 Lattice size& Previous algorithm & Present algorithm \\
 \hline
 6$\times$6$\times$6  & 200 days & 18 hours \\
 8$\times$8$\times$8 & 16 years & 10 days \\
 \hline
 \end{tabular}
\end{center}
\caption{Benchmark result on Pentium II 350MHz computer,
for the previous (full diagonalization) algorithm calculated on single node,
and the present (moment expansion) algorithm on 64 parallel nodes.
CPU time for 10,000 MC steps are shown.}
\end{table}

In Table 1 we show the benchmark results for 
 a comparison between the full diagonalization  
algorithm
and the moment expansion algorithm.
We take 10,000 MC steps
which is typically a minimum
number necessary for  accurate calculations.
Calculations based on the
full diagonalization algorithm are performed on one node
since it is difficult to make an efficient parallel computation.
64 CPUs are used in parallel for the moment expansion algorithm.
As a result, we see a large  improvement of the computational
speed by the moment expansion algorithm.
MC calculation on a $8^3$ lattice, which has been virtually impossible 
by the full diagonalization algorithm, now
turns out to be within our reach by the moment expansion algorithm.
Using the commodity PC clusters with
over 100 CPUs, it is practical to
perform a MC calculation for a system with $N_{\rm dim} \sim 10^3$.

Thus it is now feasible to 
investigate finite size clusters with systematic
series of lattice sizes.
Extrapolations to thermodynamic limits
as well as finite-size scalings
for various thermodynamical properties
will be reported elsewhere \cite{Motome00,Motome01x}.

\begin{figure}[htb]
 \epsfxsize=12cm
 \hfil\epsfbox{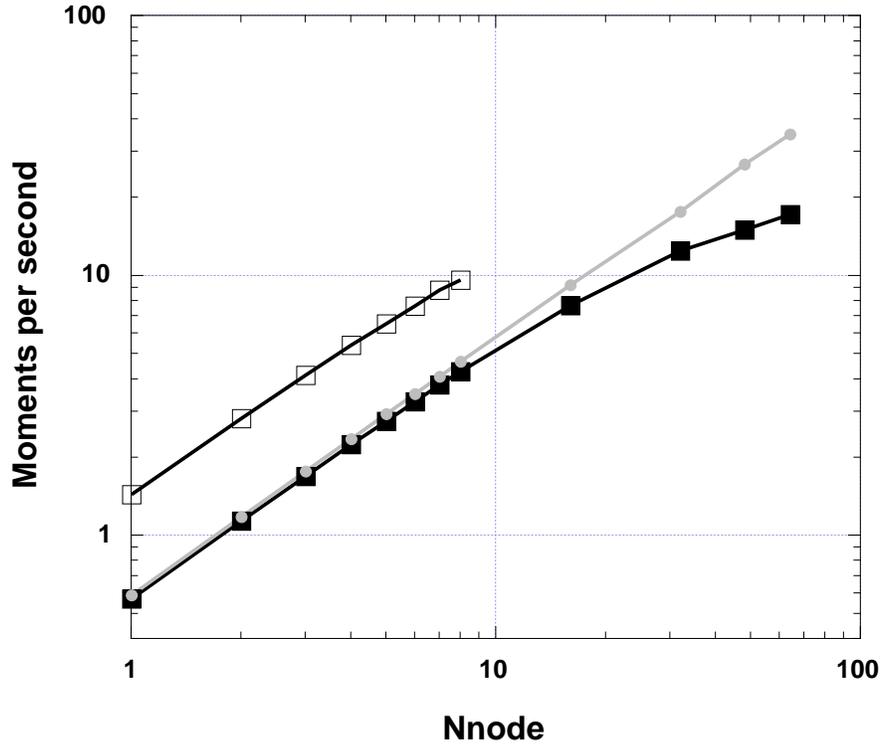}\hfil

\caption{Multiple node benchmark result of  the
 moment expansion algorithm for the double exchange model
 on a $32\times 32$ lattice.
 Calculations are performed on 
 dual PentiumII 350MHz cluster system (filled square) and
 dual Celeron 533MHz cluster system (open square)
 as well as on
 SGI2800 (filled circle in grey) as a comparison.}
\end{figure}

Benchmark results for the parallel calculation efficiency are
shown in Fig.~1. 
Computational speeds are indexed by numbers of moments
calculated per second. We show the results 
for (i) a PentiumII system, (ii)  a Celeron system, as well as
(iii) SGI2800 for comparison.
We see that in either system the computational speed
scales almost linearly with the number of nodes $N_{\rm node}$.
This indicates that the efficiency of the parallel computation
is quite high at $N_{\rm node} \simle 10^2$.
The benchmark result also shows that the program
runs equivalently or even faster on commodity personal computer
systems compared to the high-performance 
(and high-cost) parallel computational system.

The moment expansion algorithm is optimized for 
parallel computations, especially on Beowulf-type commodity systems,
in the following sense.
(i) The most CPU time consuming part of the MC run
is the calculation of $P(\{ \vec S_i\})$, which is completely
parallelized. Data transfer occurs only once for each
calculation of $P(\{ \vec S_i\})$, while the
CPU time for it scales as $O(N_{\rm dim}^2/N_{\rm node})$. Thus,
for sufficiently large systems, data transfer time is
negligibly small compared to calculation time.
High efficiency of the parallel computation is well understood by 
Amdahl's law.
(ii) In general, MC calculations consume small amount of memory.
In the present case with $N_{\rm dim}\sim 10^3$, less than 100kB of memory is
used for the vectors $|\nu;m\rangle$ as well as for the matrix $\cal H$, which
do not overflow from  the Level2 (L2) cache of commodity CPUs.
Computational speed scales almost linearly with
CPU clock speed. 
(iii) Communications among CPUs as well as access to main memories
are, in general, the bottlenecks in the usage of the commodity-type clusters.
In our algorithm, these two features conceal the disadvantages.

Since the increase of the CPU clock speed 
for commodity computers
is quite large nowadays, it is now getting easier and easier
to construct a commodity computer systems
which exhibits a high price-performance for the moment
expansion calculations.

Let us finally note that 
the moment expansion algorithm is applicable to
many kinds of  strongly correlated electron systems.
We have obtained a powerful
algorithm to investigate
thermodynamic properties of various electronic models
which have not yet been studied numerically in a
systematic way.

\section*{Acknowledgements}

The work is supported by the
Scientific and Engineering Research Center,
Aoyama Gakuin University.
MC calculations are performed 
on {\em Aoyama Plus} systems \cite{AoyamaPlus-} as well as at the
Supercomputor Center, ISSP, University of Tokyo.

\end{document}